\documentclass[10pt,twocolumn,floatfix,superscriptaddress,reprint]{revtex4-1}

\usepackage[paperwidth=210mm,paperheight=297mm,centering,hmargin=2cm,vmargin=2.5cm]{geometry}

\usepackage[pdftex]{graphicx}
\usepackage{amssymb,amsmath} 
\usepackage[utf8]{inputenc}
\usepackage[OT1]{fontenc} 
\usepackage{hyperref}
\usepackage{xcolor}

\begin{document}

\title{Headphones on the wire\\
  \vspace*{0.1cm}
  \small Statistical patterns of music listening practices}

\author{Thomas Louail}
\thanks{Correspondence and requests for materials should be
  addressed to TL (email: thomas.louail@cnrs.fr)}
\affiliation{CNRS, UMR 8504 Géographie-cités, 13 rue du four, FR-75006 Paris}
\affiliation{Instituto de F{\'i}sica Interdisciplinar y Sistemas Complejos IFISC
  (CSIC-UIB), Campus UIB, ES-07122 Palma de Mallorca}

\author{Marc Barthelemy} 
\affiliation{Institut de Physique Th\'{e}orique, CEA, CNRS-URA 2306, FR-91191
  Gif-sur-Yvette}
\affiliation{CAMS (CNRS/EHESS) 190-198, avenue de France, FR-75244 Paris Cedex 13}

\newcommand{\marc}[1]{\textcolor{green}{\textbf{@@@ Marc: #1 @@@}}}
\newcommand{\thomas}[1]{\textcolor{blue}{\textbf{@@@ Thomas: #1 @@@}}}



\begin{abstract}
  We analyze a dataset providing the complete information on the effective plays of thousands
  of music listeners during several months. Our analysis confirms a number of properties
  previously highlighted by research based on interviews and questionnaires, but also uncover
  new statistical patterns, both at the individual and collective levels. In particular, we
  show that individuals follow common listening rhythms characterized by the same fluctuations,
  alternating heavy and light listening periods, and can be classified in four groups of
  similar sizes according to their temporal habits \-- ``early birds'', ``working hours
  listeners'', ``evening listeners'' and ``night owls''. We provide a detailed radioscopy of
  the listeners' interplay between repeated listening and discovery of new content. We show
  that different genres encourage different listening habits, from Classical or Jazz music with
  a more balanced listening among different songs, to Hip Hop and Dance with a more
  heterogeneous distribution of plays. Finally, we provide measures of how distant people are
  from each other in terms of common songs. In particular, we show that the number of songs $S$
  a DJ should play to a random audience of size $N$ such that everyone hears at least one song
  he/she currently listens to, is of the form $S\sim N^\alpha$ where the exponent depends on the
  music genre and is in the range $[0.5,0.8]$. More generally, our results show that the recent
  access to virtually infinite catalogs of songs does not promote exploration for novelty, but
  that most users favor repetition of the same songs.
\end{abstract}

\maketitle

The reasons why human beings like listening to music, the variety of emotions music
can arouse, its uses and functions in human societies: those are some long lasting questions
which have been discussed by music critics and by scientists belonging to a wide range of
disciplines. From the early musicology \cite{Adorno:1938} to popular music studies
\cite{LeGuern:2010} through sociology of cultural practices \cite{Bourdieu:1984}, geography
\cite{Carney:1998, Lee:2012}, music history \cite{Cohn:1972, Guralnick:1986}, cultural
economics \cite{Cox:1995,Prieto-Rodriguez:2000}, educational and cognitive psychology
\cite{Sloboda:1999, North:2000, Miranda:2009, Hallam:2011}, physiology and neurosciences
\cite{Levitin:2006, Salimpoor:2009}, an eclectic scientific litterature has illuminated many
different facets of music listening. At a collective level, it has been demonstrated several
times that statistical relations between inherited social characteristics of individuals and
their musical preferences exist \cite{Bourdieu:1984, Prior:2011, Bryson:1996,
  VanEijck:2001}. At the individual level, studies relying on questionnaires, interviews or
experiments conducted in controled environments have documented both the functions attributed
to listening and the emotions aroused, in various situations of daily life and in different
contexts \cite{Juslin:2004, Sloboda:1999, Hallam:2011, Salimpoor:2009}. The influence of the
device on the listening practice \cite{Krause:2013}, the effects of listening on a
number of daily activities -- e.g. performance at work \cite{Lesiuk:2005}, driving
\cite{Dibben:2007}, coping and regulating emotions \cite{Miranda:2009, Saarikallio:2010} -- or
the rewarding aspects of music-evoked sadness \cite{Taruffi:2014} are other examples of
listening-related research. Classifications of listeners have been proposed, with some authors
concluding about the existence of a direct relation between musical preferences and cognitive
styles \cite{Greenberg:2015}, other stressing the uses of music and their self-declared
importance as relevant classifiers \cite{TerBogt:2011, Schafer:2016}. Statistical physicists
also contributed by highlighting structural properties of artists and genres communities that
emerge from the analysis of personal libraries of audio files \cite{Lambiotte:2005}.

However, relatively little is known about \emph{how precisely} we listen to recorded music on a
daily basis. By \emph{how} we refer here to some kind of detailed, quantified radioscopy of our
contemporary listening practices of recorded music, an important aspect of the relation we
entertain with music.

Until recently, any empirical research willing to answer to questions pertaining to daily
listening practices had to rely on surveys and interviews. The technological and societal
evolutions have sustained the development of new mobile devices, online tools and listening
possibilities, as well as new actors in the music industry. Music-on-demand services have
quickly gained in popularity over the last few years, and for example, according to a recent
report of the French national syndicate of phonographic publishing \cite{SNEP:2016}, more than
three million of French residents (approx. 4\% of the total population) were subscribing to an
on-demand streaming music platform in 2016, and roughly $1/3$ of the total french population
regularly stream audio content. The data recorded by 
streaming platforms offer great possibilities to 
analyze and hopefully better understand individual and collective listening practices.

Whenever an individual plays a song through such a service, with a web browser or dedicated
application, online or offline, all known information associated with the stream are logged in
the company's database. For example, when Amélie plays The Roots' \emph{Kool On} on her mobile
phone, a new entry is added to the database, informing that at 8:23 am on 2/10/2015 she played
the song (\emph{Kool On}) by \emph{The Roots}, from the album \emph{Undun}. Incidentally, we
also know the genres tags associated to the song, the stream's duration (and consequently if
she listened to the song entirely or not), the information declared during registration,
including her age and city of residence, and possibly some additional contextual information
(e.g. if the song was part of one of Am\'elie's playlists, if she was online or offline when
listening to the song, possibly the city where Am\'elie was located when listening, etc.). Such
a record of information is added to the database for each single 
play of the millions of registered users of the service. Once anonymized, users' listening
history data can be processed and analyzed, and those digital traces passively produced while
listening consequently constitute an unprecedented empirical source to study quantitatively
contemporary listening practices.

In the following we analyze the listening history data of one of the major streaming platforms.
The data correspond to the entire listening history of about five thousands users during one
hundred days (see Material for details).  We note that for some of these people the music
streamed
may represent only a limited subset of all the recorded music they listened to during that
period, and their data might not be representative of their entire listening practice
\cite{Krause:2013}. In order to control this bias we selected a set of anonymous users among
those who displayed a frequent use of the service (see Appendix). For these
listeners we can reasonably assume that the streaming platform, if not exclusive, constitutes a
daily source of music.

\section*{Results}

\subsection*{Rhythms}

\begin{figure*}
  \centering
  \includegraphics[width=\textwidth]{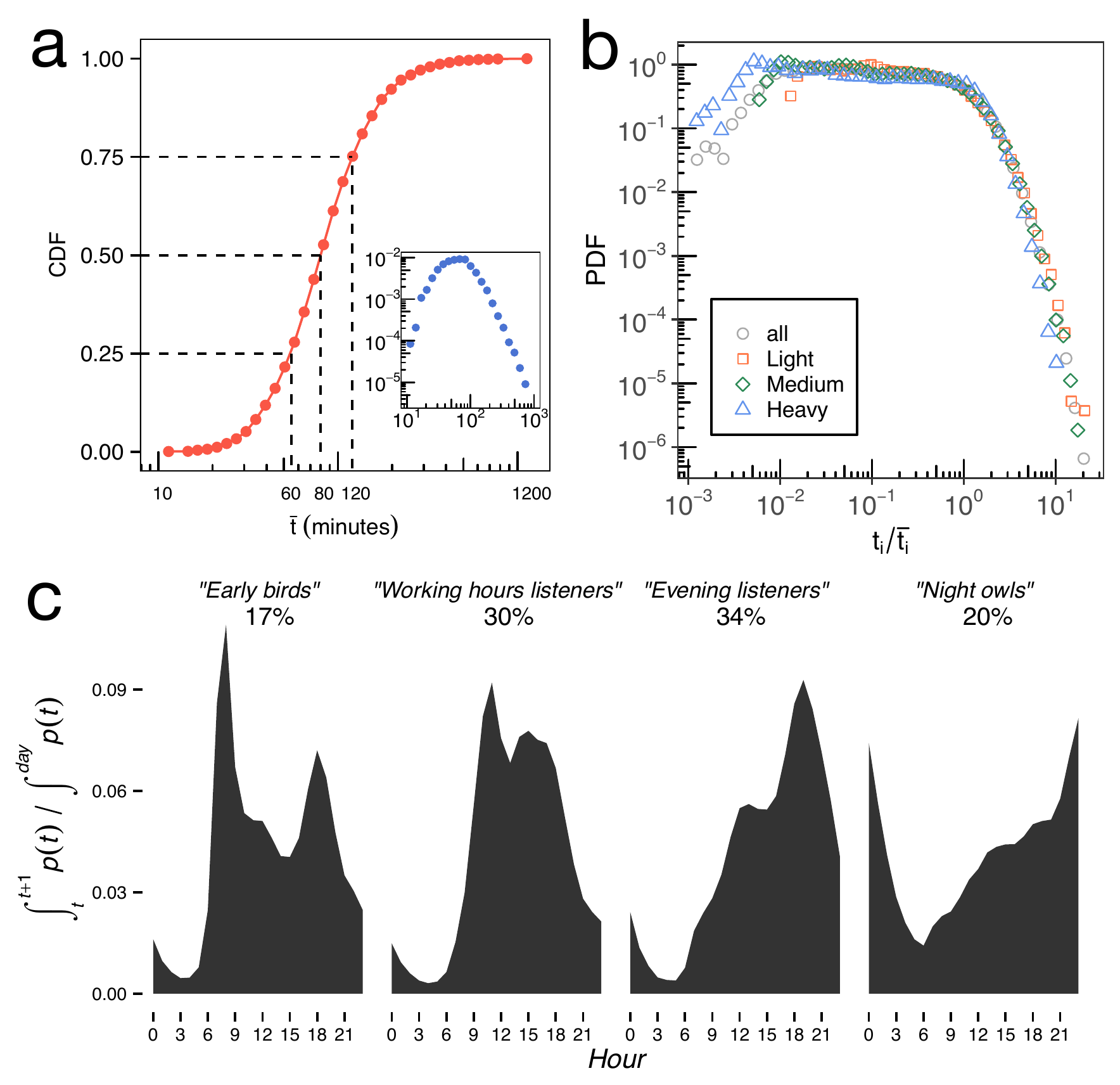}
  \caption{\label{fig:time-properties} \textbf{Music listening rhythms.} (A) Cumulative
    distribution of the average daily listening time $\bar{t}$ (in minutes) computed over all
    listeners and all day. The y-value of each point gives the fraction of users that listened
    at most $\bar{t}$ minutes of music each day on average, during the 100 days period under
    study -- inset: corresponding probability distribution. (B) The normalized distributions of
    daily listening times $(t_i/\bar{t_i})$ for three different groups of listeners: heavy
    listeners (more than 2 hours per day on average, blue triangles), medium (between 1 and 2
    hours, green diamonds), and light (less than 1 hour per day, orange squares).
    (C) Average profiles of different groups of listeners obtained by clustering on their
    listening temporal habits. 
  }
\end{figure*}

We start by quantifying the relation that individuals have with music listening as a daily
activity, and the rhythms and typical hours at which they perform this activity. For all
individuals, we compute the total time they spent listening 
during the 100 days period under study. 
Fig.~\ref{fig:time-properties}a represents the cumulative distribution of the average daily
listening time $\bar{t}$ computed over all individuals and all days. This quantity varies
theoretically from $0$ to $1440$ (number of minutes in a day) and the empirical measure reveals
a range from $4$ to $1200$ minutes per day, a median value of approximately $80$ minutes, and
about $75\%$ of individuals listening more that one hour per day (and $25\%$ listening more
than 2 hours per day). In order to understand how listeners behave over 
periods of several months, we extract the distribution of the daily listening times $t_i$ for
all individuals (the index $i$ refers to the individual) and for the whole period, splitting
listeners into three groups according to their total listening time (``light'', ``medium'' and
``heavy'' listeners). We compute for each individual its average $\bar{t_i}$ over all days and
we show on Fig.~\ref{fig:time-properties}b the normalized $(t_i/\bar{t_i})$ distributions for
the three groups. These normalized distributions collapse onto a single curve, indicating that
whatever how much music they listen to, individuals display a common behavior characterized by
the same fluctuations in listening times, alternating days with relatively few music listened
and days of heavy listening. These distribution are peaked and can be fitted by an exponential
function, suggesting a Poisson nature of the listening behavior, in contrast with previous
results on daily human behavior \cite{Barabasi:2005}.

We then wonder at what time of the day people listen to music. We introduce $u(h)$ which counts
the number of individuals listening to music at time $h$. In order to highlight the collective
rhythm we plot the normalized values $\int_h^{h+1} u(h) / \int_0^{23} u(h)$, where
$\int_0^{23} u(h)$ is the total number of unique individuals that listened to music during the
day (see Supplementary Figure 1). Like other daily activities which have been heavily studied
from individual traces \cite{Lenormand:2014,Lenormand:2015} the aggregated curves
of activity display two characteristic patterns, one for weekdays and another for
weekends. 
However 
not everyone listens to music at the same hours, and for each listener $i$ we calculate his/her
proportion $p_i^h$ of plays that occurred between $h$ and $h+1$, averaged over the 100 days
period. We then construct the 24-values vector $(p_i^0, \dots, p_i^{23})$ and using these time
profiles we cluster the listeners, and find four typical groups whose average profiles are shown
in Fig.~\ref{fig:time-properties}c (see the Appendix for details on the clustering). These time profiles
are very specific: in one group individuals listen to music mostly in the morning (``Early
birds''); in another we observe two listening peaks, one in the morning and the other in the
afternoon (``Working hours listeners''); there are also individuals listening to music mostly
at the end of the day (``Evening listeners''); and finally those whose listening peak is late
in the evening and during the night (``Night owls'').

\subsection*{Difference and repetition}
\begin{figure*}
  \centering
  \includegraphics[width=\textwidth]{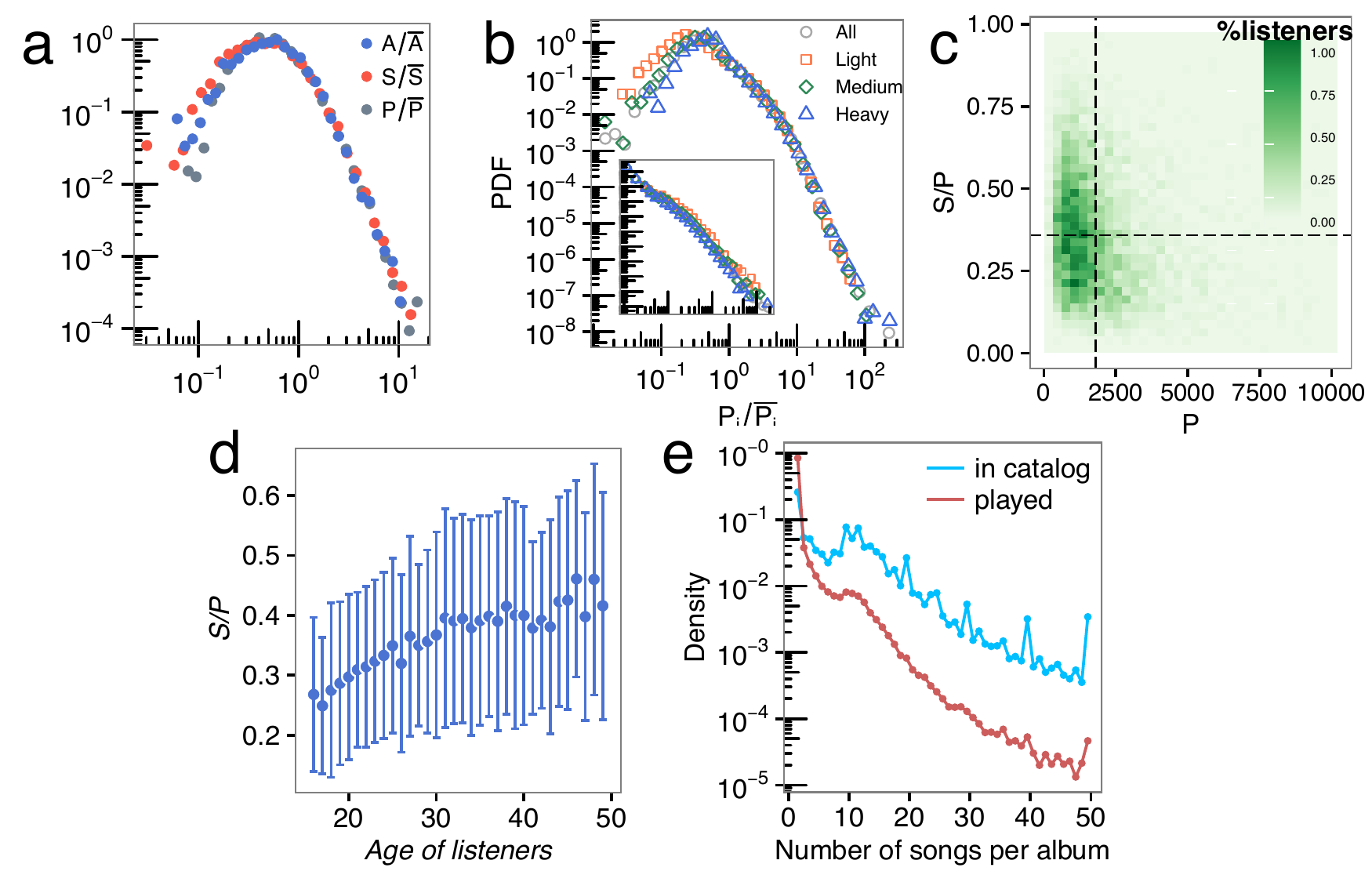}
  \caption{\label{fig:typology}\textbf{Dimensions of contemporary listening practices.}  (A)
    Histogram of the total numbers of plays ($P$), songs ($S$) and artists ($A$) listened among
    users during the $100$ days period. (B) Normalized distributions of plays per song
    $P/\bar{P}$ for three groups of listeners (light, medium and heavy -- according to their
    total number of plays). Inset: distribution $p(P_i(\alpha))$ for the three groups of
    listeners. (C) Binned scatterplot of the listeners' exploratory ratio $S/P$ vs. their total
    number of plays $P$. (D) Exploratory ratio $S/P$ vs. listeners'age. For each age, the error
    bars correspond to a standard deviation. 
    (E) Histograms of the number of songs per album in the music catalog available to listeners
    (blue curve) and of the number of songs \emph{played} per album listened (red curve).}
\end{figure*}

We now investigate how individuals are distributed along various dimensions of music
listening. 
We denote by $P_i$, $S_i$, and $A_i$ the total numbers of plays, unique songs and unique
artists listened by individual $i$ during the 100 days period, respectively. We show on
Fig.~\ref{fig:typology}a the distribution $p(P/\bar{P})$, $p(S/\bar{S})$ and $p(A/\bar{A})$ of
the normalized variables (where the average values are computed over all individuals and the
whole 100 days period). These distributions collapse on a curve that can be fitted by a
log-normal distribution of parameters $\mu\approx 3.9$ and $\sigma\approx 1.1$. It is here
another 
occurrence of the lognormal distribution in social dynamics, although its origin here is not
clear and would deserve further investigation. Also, while we can understand a priori that $P$
and $S$ display the same behavior, it is more surprising that the distribution of the number of
unique artists listened also collapses on the same curve.

In order to understand how individuals distribute their plays among the songs they listen to,
we compute the aggregated distributions $p(P_i(\alpha))$ which contain the numbers of plays per
song $\alpha$ for each listener $i$ (we then have $\sum_\alpha P_i(\alpha) = P_i$) for the
three groups defined above -- heavy, medium and light listeners, according to their total
number of plays. We represent in Fig.~\ref{fig:typology}b these distributions
$p(P_i(\alpha)/\overline{P_i(\alpha)})$ (with $\overline{P_i(\alpha)} = P_i/S_i$) which also
collapse on a curve whose tail can be fitted by a power law with an exponent $\approx 3.0$ (the
distributions $p(P_i(\alpha))$ are shown in the inset). The fluctuations around the average
value $P/S$ are therefore the same whatever the group: no matter how much music people listen
to, they distribute similarly their attention on the different songs listened. Patterns of
Fig. \ref{fig:typology}a and b might result from a simple relation between $P$, $S$ and $A$
common to all listeners and that would allow to make predictions for any of these variables
knowing the value of another. As shown on Supplementary Figure 4, there are however no clear
relations among these variables and we observe large fluctuations from one person to another.

We now focus on the quantity $S_i/P_i$ which can be seen as the exploration rate of individual
$i$ among the catalog. By definition this ratio varies between 0 -- the extreme case of an
individual that would listen to one and only song again and again -- and 1 -- a pure explorer
that would never listen a song twice. The binned scatterplot of Fig.~\ref{fig:typology}c shows
that there is no clear relation betweeen the weight of exploration $S/P$ and the total time
spent listening to music (characterized by $P$). We also see that the average value of $S/P$ is
around $1/3$, indicating that the average number of plays per song is $\approx 3$. We observe a
trend (despite large fluctuations) between the average rate $\overline{S_i/P_i}$ and the
listeners' age (shown in Fig.~\ref{fig:typology}d; errors bars correspond to one standard
deviation), indicating a decrease of repetition and an increase of exploration with age (see
also \cite{Lamere:2014a}).

Streaming audio is a different experience than listening to the radio or browsing in a
personnal collection of records or audio files. Several modes are offered to users: they can
search and play songs one after another, listen to an entire album or listen to a playlist
previously compiled by themselves, someone else or automatically generated (the latter gaining
in importance thanks to increasingly sophisticated recommendation systems
\cite{Bonnin:2014}). Vinyle records favor a sequential listening from the first to the last
track, CDs allowed direct access to any song of the record but still contain albums which are
``meant to'' be played entirely. Streaming platforms offer listeners an immediate access to any
song. The possibility to pick songs among a practically infinite catalog suggests the naive
assumption that we should observe a high versatility in plays, and listening sessions mixing
album-centered habits with handmade sequences of songs (in playlists or not). In order to check
this hypothesis we first compute for each listener the percentage of albums listened entirely
(while not necessarily in sequential order), and plot the distribution of this percentage among
the population of listeners on Supplementary Figure 5. 
More than $50\%$ of users played in their entirety less than $5\%$ of the albums. 
We then compute the distribution of the number of songs \emph{played} per album, and compare it
to the distribution of the number of songs per album. If listeners had album-centered
practices, then both distributions should match. The results shown on Fig.~\ref{fig:typology}e
tell a different story. The distribution of the number of songs per album in the catalog (blue
curve) has several small peaks around typical values that correspond to different types of
records: 1 (singles), 10--12 (the typical number of songs on albums), and then 20/30/40 (likely
corresponding to double/triple albums and compilations/anthologies). In contrast we observe a
very different distribution (red curve) when we consider the number of songs played per album,
that displays a regular decay with a smaller peak around 10, corresponding to the remainder of
album-centered listening practices.

\subsection*{Music genres and listening habits}

Each song is indexed with one or several genre tags. While there are hundreds of unique tags in
such songs databases, most of them are associated to a very small proportion of songs only, and
concern an even smaller proportion of plays. The distribution of the listeners' plays per genre
shows us first that over a period of several months, most individuals listen to songs of very
different genres (see the histogram of the number of genres listened at least once on
Supplementary Figure 8). This first impression of broad eclecticism is challenged by a closer
look at the individuals' distributions of plays among genres. For each individual $i$ we
compute the Gini normalized coefficient (see Supplementary Methods in Appendix) of his/her
distribution $p(P_g(i))$ of plays in each music genre $g$, and plot on Supplementary Figure 9A
the distribution of this Gini coefficient among listeners. We first observe that there are no
listeners displaying small Gini values. On the contrary, most individuals have large Gini
values, indicating that even eclectic individuals who listen to many different genres tend to
strongly favor a subset of them. From the value of the Gini coefficient we extract a typical
number of dominant genres among the listener's plays (see Supplementary Methods in
Appendix). It appears that most individuals have 2 or 3 genres that they clearly favor
(cf. Supplementary Figure 9B and
C). 
This observation naturally leads us to determine the couples of genres which often go together.
We then determine for each listener his/her two most listened genres, and estimate the
probability $P(g_2|g_1)$ to have $g_2$ as second favorite genre when $g_1$ is the favorite
one. These probabilities are represented on Supplementary Figure 10. Beyond the leading role of
Pop music on streaming platforms, we recognize classic proximities, such as Metal-Rock, the Hip
Hop family or the Classical-Jazz tandem. We also observed that the temporal patterns of the
music genres do not strongly differ from each other (see Supplementary Figure 11)
\cite{Jagger:1969}. Similarly, we could not distinguish groups of artists that are
preferentially listen to at certain hours \cite{Sinatra:1955}.

\begin{figure*}
  \centering
  \includegraphics[width=\textwidth]{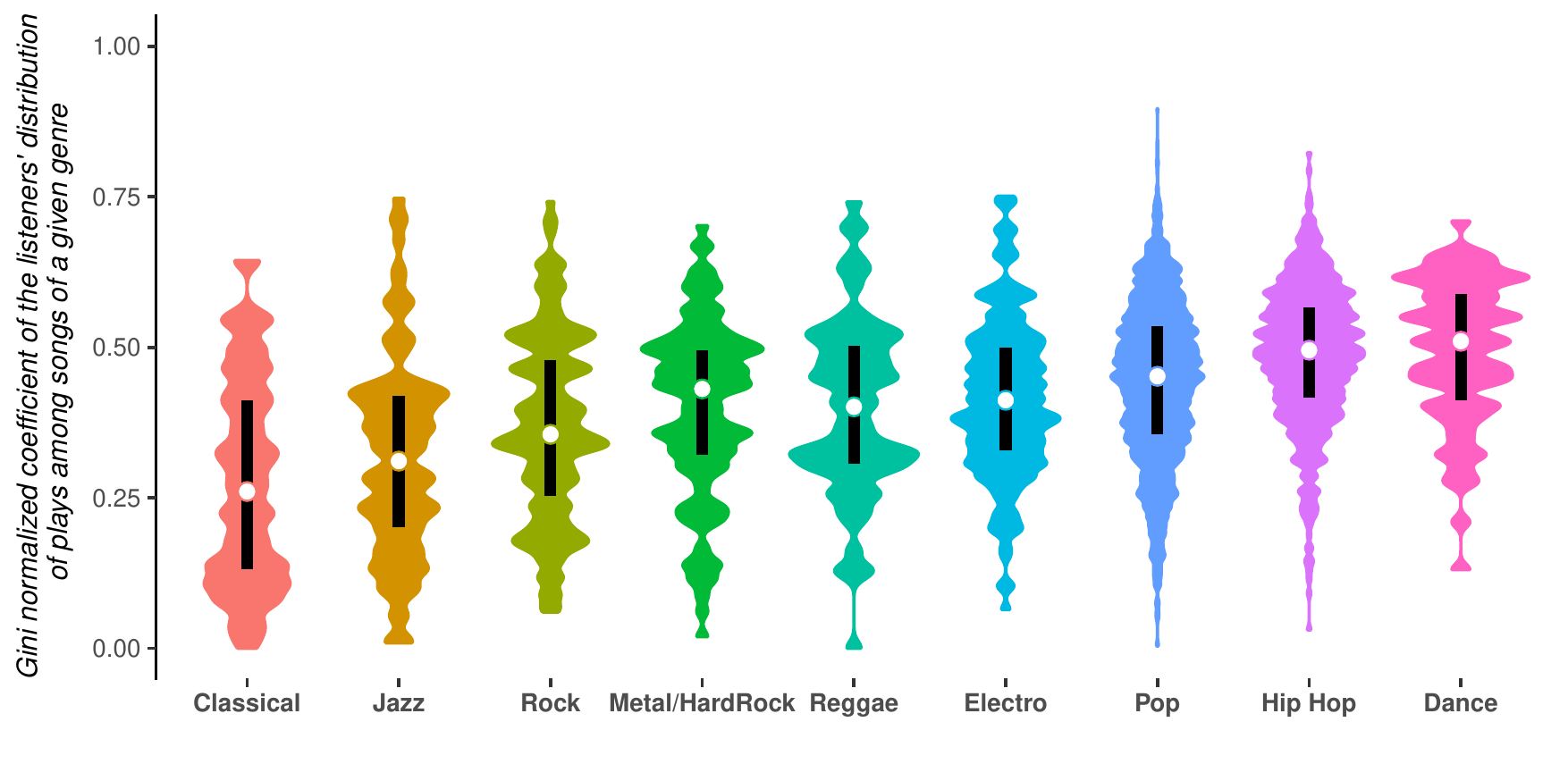}
  \caption{\label{fig:genres-panel} \textbf{Inequality of music genres.} Each violin is a
    vertical and symetric representation of the smoothed distribution (kernel density
    estimation) of its listeners' Gini coefficient of plays per song, for each genre. Each
    violin shows to what extent individuals have an heterogeneous attention towards the songs
    they listen to, when they listen to that genre.}
\end{figure*}

We now focus on 9 broad genres (Classical music, Jazz, Rock, Metal/Hard Rock, Reggae, Electro,
Pop, Hip Hop, and Dance) to see if this classic, ``record store alleys'' classification allows
to discern various listening habits. Considering how recorded music is produced and
distributed, some music genres favor the emergence of ``hits'' and are more ``inegalitarian''
than others when it comes to the repartition of the crowd's attention towards songs (see
Supplementary Figure 12). The heterogeneity of the number of plays in each genre is represented
by a ``violin plot'' shown in Fig.~\ref{fig:genres-panel}. Each genre is represented by a
violin, which gives vertically and symetrically the smoothed distribution (kernel density
estimation) of the listeners' Gini coefficient for songs in that genre. The Gini coefficient of
a given individual for a given genre encodes the inequality of his/her distribution of plays
among the songs of this particular genre. The violins are shown from left to right according to
the average value of the individuals' Gini coefficients. These violins show that the Gini
coefficient is usually distributed over the whole range $[0,1]$, indicating that for most
genres there is a strong heterogeneity of listening practices. We however observe that
individuals, when they listen to the most popular genres (Pop, Hip Hop or Dance), are more
homogeneous and seem more focused on a subset of songs. In contrast, in other genres we observe
a greater variety of listening practices (for example for Jazz or Classical). We then group
listeners according to their favorite genre: Rock, Rap, Jazz, etc. We calculate the average age
of the people in each group, along with the average exploration rates $S/P$ inside and outside
the favorite genre (shown on the Supplementary Figure 13). We note that in all groups the
exploration rate is larger outside the favorite genre than inside, an indication that for most
individuals what contributes to make a genre their favorite is the repetitive listening of a
small number songs of that genre. We also note substantial differences in terms of the average
age of listeners in the different groups, an expected observation in agreement with previous
work \cite{LeBlanc:1999,Prieto-Rodriguez:2000, Lamere:2014a}.

\subsection*{Playing music at a party}

\begin{figure*}
  \centering
  \includegraphics[width=\textwidth]{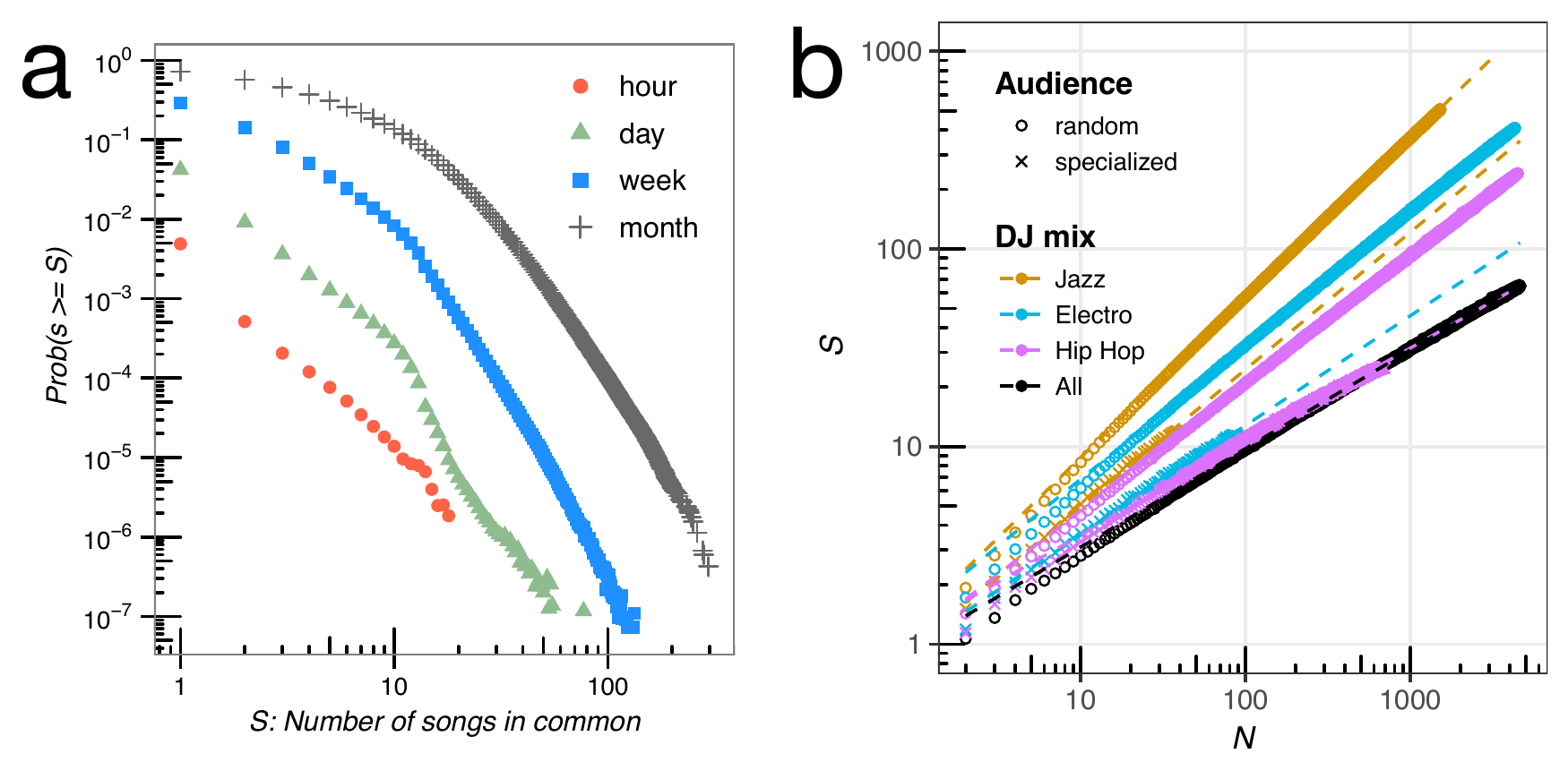}
  \caption{\label{fig:assembly} \textbf{Organizing a party.} (A) Estimated probabilities that
    two randomly chosen individuals listened at least to $S$ common songs, for different
    periods of time. (B) DJs' functions capturing the relation between the size $N$ of an
    audience and the minimal number of songs $S$ one should play so that everybody in the room
    hears at least one song that he/she recently listened to. Colours correspond to DJs playing
    specific music genres, while shape of points correspond to random or specialized
    audiences.}
\end{figure*}

Finally, we provide measures about the ``distance'' between individuals in terms of the music
they listen to. The size of the online music library is practically infinite which implies that
individuals' plays would have a very small overlap if random. Musical choices however depend on
many things, are strongly influenced by social factors \cite{Bourdieu:1984} and we could expect
a larger value of the overlap than the one obtained by chance. We first estimate from the
dataset the probability that two randomly selected individuals share a given number of
songs. Fig.~\ref{fig:assembly}a shows the probabilities $p=Prob(s\geq S, \delta t)$ that they
shared at least $S$ songs during a period $\delta t$ varying between one hour and one
month. This probability obviously increases as one considers longer time periods, and the
probability $p(S\geq 1, \delta t=30\;days)$ that two people have listened at least one common
song in 30 consecutive days is large $p \geq 0.7$ ($p \geq 0.9$ for a 100 days period). The
four curves on Fig.~\ref{fig:assembly}a are very similar and display a cut-off value of
$\approx 10-15$ songs.

A related problem is to determine the minimum number of songs that a DJ should play to an
audience so that everyone hears at least one song that he/she recently listened to. This
problem is well-known by non-professional DJs who have to deal with people harassing them to
play specific songs at parties. This minimum number of songs $S$ obviously depends of the size
$N$ of the audience, and we call the function that relates $S$ to $N$ the \emph{DJ
  function}. In order to evaluate empirically this function, we randomly sample different sets
of listeners of increasing size $N$, and for each set we determine the smallest number of songs
$S$ that allow to satisfy them all (in other words, $S$ is the size of the minimum 1-mode
vertex cover in the bipartite subgraph connecting the individuals sampled and the songs they
listened). When plotting this number $S$ vs. $N$, we observe a behavior of the form
$S \sim \sqrt{N}$, as shown by the black curve on Fig.~\ref{fig:assembly}b. We repeat the same
calculation by focusing on specific music genres to cover the case of ``specialized'' DJs. We
select songs (and their listeners) of a given genre only, allowing us to evaluate a DJ function
per genre. Each of them has the same general form $S \sim N^\alpha$, with
$0.64 \leq \alpha \leq 0.8$ ($R^2 > 0.99$). These exponents give a lower bound to the size of
the setlist that one has to play if she/he wants to ``satisfy'' everybody in a random crowd
full of strangers. In particular, if the venue is big and the audience large ($\geq ~10^{4}$
individuals), the required number of songs will be too large to be played during a single
event, making the challenge impossible whatever the DJ. In reality, the crowd attending to a
gig is not random and gather individuals with similar taste. We then reproduce the same
experiment but this time by considering specialized audiences, composed of people whose
favorite genre is the one played by the DJ. As expected we obtain smaller exponents, with
$0.47 \leq \alpha \leq 0.8 $ ($R^2 > 0.99$), showing that specialized audiences are easier to
``satisfy''.

\section*{Discussion}

Streaming platforms contribute to increase possibilities of access to recorded music, and might
possibly change the listening habits of their users. 
These can experience legal and unlimited access to a gigantic catalogue containing more years
of unique music than what one could listen to in an entire life. Our results challenge a number
of naïve assumptions about the contemporary forms of an old and widespread cultural
practice. 
Considering the available catalog, one could think that it
would 
encourage listeners to continuously search for novelty, and browse in many genres, artists and
albums. Our results on the weight of repetition and the small number of dominant genres per
listener indicate that it is currently not the case. 
This observation takes place in the context of discussions about the psychological function of
repetition when listening to music. These considerations are however beyond the scope of this
article, and we refer the readers to more specific work relying either on detailed interviews
with listeners \cite{Sloboda:1999,Kamalzadeh:2016}, self-reports \cite{TerBogt:2011} or
neuroimaging \cite{Levitin:2006}. We remind however that ``talk is cheap''
\cite{Jerolmack:2014}, and our observations that (i) heavy and light listening days identically
alternate whatever the perceived importance of music in the listeners' daily lives, and (ii)
that the weight of repetition is independent of the amount of music listened, challenge
previous results based on questionnaires and interviews.

It would be wise nonetheless not to generalize too hastily our results to music listening
practices in general, whatever the listening device. The availability of pre-existing playlists
and the automated generation of playlists fitted to the users' taste might encourage a less
involved listening process, resulting in distinct statistical properties between ``active'' and
``passive'' listeners. For example, the authors of \cite{Greasley:2011} concluded that those
who declare listening more music are also more involved in the choice of the music they listen
to. The data we analyzed include no contextual information that let us know if the songs
listened were voluntarily played after a proper search, or if they were recommended and queued
by the service itself. We should also mention that there is so far a limited proportion of
individuals who use streaming platforms as their main source of recorded music, but this
proportion is constantly increasing \cite{SNEP:2016}. 
We have restricted our analysis to listeners whose activity suggests that they favor
streaming. But considering that these may not representative of the entire population (in terms
of social and demographic criteria), our results need to be confirmed with richer datasets
providing more contextual information.

The collection of individual data by companies raises privacy issues and legitimate concerns
about surveillance. For obvious reasons these companies are reluctant to share the raw data
they collect, even after proper anonymisation. This policy partly explains why very few results
obtained from such data have been published so far in the scientific literature. Questions
similar to those we addressed are nonetheless studied internally in a product-oriented research
(see for example \emph{Spotify Insights} \cite{SpotifyInsights} or \emph{Music Machinery,},
collections of blog posts discussing data-driven analysis of listening practices, e.g
\cite{Lamere:2014}). However, ``digital footprints'' alone do not give researchers clues about
the individuals' intentions explaining their behavior and choices. More generally, these traces
poorly inform about the context of use, suffer several uncontrolled bias, and might lead to
misinterpretating the results \cite{Lewis:2015} (e.g. was the user really listening -- or even
in the room -- when the song was played?). Consequently any ``blind'' analysis of logs alone is
doomed to be limited in scope, and in some cases may lead to wrong conclusions (e.g. see
\cite{Gallotti:2016} for a discussion of the case of individual human trajectories
reconstructed from unconventional data sources). While an increasing part of daily human
activities produce electronic traces, designing information collection protocols which
articulate the strengths of both traditions (detailed surveys and interviews in one hand and
digital footprints in the other) is a contemporary challenge faced by social research. 

\section*{Material and Methods}

\subsection*{Dataset}
The dataset analyzed contains the raw streaming data of 10,000 anonymous and registered
users of a major streaming platform. They inform us on their entire listening history during
the 6-months period spanning from June 1st 2013 till December 1st 2013. These users were
randomly selected among all the registered French users of the service. We know nothing about
how important is the streaming service for these users, who might also heavily rely on other
music sources and devices (their own personal library of records or audio files, the radio,
etc.), and might have distinct practices depending on the source and device \cite{Sloboda:1999,
  Krause:2013, Kamalzadeh:2016}. In order to mitigate this bias, we chose to focus on users who
made a frequent use of their account. We selected a set of 4,615 anonymised French users who
actively listened to music (at least every other day in average) during a 100 days period, from
2013/8/15 to 2013/11/23. For these individuals we can reasonably make the hypothesis that
streaming, if not their unique, was one of their main music source during this period (see
the Appendix for details on the cleaning and filtering of the data).

\subsection*{Normalized Gini coefficient and extraction of the
  number of dominant terms}
We assume that we have $K$ classes and in each class, we have a random number $X_i$. The Gini
coefficient can then be computed as follows
\begin{equation}
G_K(X)=\frac{1}{2K^2\overline{X}}\sum_{p,q}|X_p-X_q|
\end{equation}
This coefficient is a priori in the interval $[0,1]$ but we will see that for finite $K$ the
maximum value is actually different from 1 and depends on $K$.

\subsubsection*{Case $K$ and $D=1$}

We denote by $D$ the number of dominant terms. If $D=1$ we have only one dominant term that we
call $X_1=a$ and all the other terms are much smaller than $a$ and for simplicity -- without
any loss of generality -- we take them $X_{i\neq 1}=1$. The Gini coefficient is then
\begin{align*}
  G_K(a)&=\frac{1}{K^2}\frac{(a-1)(K-1)}{\frac{a+K-1}{K}}\\
        &=\frac{K-1}{K}\frac{a-1}{a-1+K}
\end{align*}
In the limit $a\gg K$ where the heterogeneity is maximal and the Gini coefficient maximum, we
then obtain
\begin{equation}
  G^*_{K}=\frac{K-1}{K}
\end{equation}
We see on this formula that $G^*$ can actually be much smaller than 1 if $K$ is not too
large. In the case of a small $K$ we thus have to compute the normalized Gini coefficient
$G/G^*_K$ which is in the interval $[0,1]$ and reaches $1$ for the most heterogeneous
distribution.

\subsubsection*{General $K$ and $D$ case}

We assume here that we have $D$ dominant terms and for simplicity we assume that
$X_1=X_2=\dots=X_D=a$ and $X_{i>D}=1$. We then have
\begin{equation}
\overline{X}=\frac{Da+(K-D)}{K}
\end{equation}
We also obtain
\begin{equation}
\sum_{p,q}|X_p-X_q|=2D(K-D)(a-1)
\end{equation}
and the Gini coefficient is then
\begin{align}
G_{K,D}(X)&=\frac{1}{2K^2}\frac{2D(K-D)(a-1)}{\frac{Da+(K-D)}{K}}\\
&=\frac{K-D}{K}\frac{D(a-1)}{Da+K-D}
\end{align}
In the limit where $a\gg 1$ we then obtain the maximum value
\begin{equation}
G^*_{K,D}=\frac{K-D}{K}
\end{equation}

\subsubsection*{Extracting the number of dominant terms}

For a given observation of the Gini coefficient for $\{X\}$, we can ask what is the equivalent
configuration with $D_{eff}$ dominant terms ?  In other words, if we measure G, this value is
bounded
\begin{equation}
\frac{K-D}{K}<G<\frac{K-D+1}{K}
\end{equation}
where the number of effective dominant terms is given by
\begin{equation}
D_{\text{eff}}=E[K(1-G)]
\end{equation}
where $E[x]$ denotes the nearest (lower) integer of $x$.

\subsection*{Data availability}
The raw data that support the findings of this study were provided by a third-party
company. Restrictions apply to the availability of these data, which were used under license
for the current study, and are not publicly available. Derived, aggregated data supporting the
findings presented in this article are however available from the corresponding author upon
request.

\section*{Acknowledgments}
The authors thank A. Sinton and A. H\'erault for supporting the research, and A. Sinton for his
help when processing the data. T.L. thanks G. Ramelet, P. Chapron, Y. Renisio, A. Beaumont,
R. Gallotti, R. Louf, J.-L. Giavitto, F. Lamanna, W.  Nowak, N. Vallet and L. P. Vallet for
inspiring discussions on the analysis of listening practices. TL gives another high five to
G. Ramelet and L. Nahassia for their feedback on an early version of the manuscript. Final
thanks go to C.A. Burnett, T. Caruana, M.D. McCready, J.N. Osterberg, A.L. Peebles and
D. Robitaille.


\onecolumngrid
\vspace*{2cm}
\newpage
\twocolumngrid

\makeatletter
\renewcommand{\figurename}{Supplementary Figure}
\renewcommand{\fnum@table}{\sf\textbf{\tablename~\textbf{S}\textbf{\thetable}}}
\makeatother

\setcounter{figure}{0}
\setcounter{table}{0}
\setcounter{equation}{0}

\section*{Appendix}

\subsection*{Data pre-processing} After signing a non-disclosure aggrement (NDA) with a major
music-on-demand company, we were given access to a dataset containing the raw streaming data of
10,000 anonymous users of the service. These anonymous users were randomly selected among all
the registered French users of the service. The streaming data correspond to their entire
listening history during the 6-months period spanning from June 1st 2013 till December 1st
2013. The users were sampled uniformly among all their French users (no bias regarding
suscription plan, listening activity, sex, age, geographical location or years of
use). Consequently the listening data are those of users with different types of usage of the
service. In particular, some of them are paying suscribers,
while some other aren't ("free" suscribing users). The latter have to listen advertisement
between songs, cannot listen offline, the audio quality is lower, and for some of them the
total listening time is bounded. We note that these combined aspects can obviously influence
the listening activity, probably decreasing the time spent using the service each day.

Our goal was to focus on individuals who use the music-on-demand streaming platform as one of
their main music sources (if not the main one). Consequently we applied a number of filters to
select relevant users and eliminate those whose streaming data may not be representative of
their listening habits in general, whatever the source and device. The data include a few
information on the users' profiles (self-declared age, sex and city of residence), but we do
not know if the user is a paying suscriber or not, and could not get this information from the
company. It prevented us to simply filter them.

From the data we inspected the number of unique users per day and realized that it displays
large fluctuations. Not all users were active during the entire 6-month period (some appearing
only after a given date, some disappearing). To circumvent this we focused on a 100 days period
(from 15/08/2013 to 22/11/2013) during which the number of unique listeners per day remained
stable. We then selected users who displayed regularity in their use of the service during this
100 days period and used the service one every two days in average. Hence the filter is not on
the total activity (listening time) of the users but on their frequency of use. We ended with
4,615 anonymous users.

\subsection*{Supplementary Methods}
\label{sec:methods}

\subsubsection*{Clustering listeners according to their daily streaming rhythms}

For each user $i$ we construct a 24-values vector $(p_i^0, \dots, p_i^{23})$, where $p_i^{h}$
is the proportion of the user's plays that took place between hour $h$ and hour $h+1$, during
the 100 days period. We clusters these vectors with the k-means method. A usual question when
clustering $n$ individuals into $k$ clusters (with $k$ fixed a priori) is to determine an
appropriate value of $k$. A small $k$ value will result is a very simple picture but which may
poorly capture the fluctuations in the data, while $k \sim n$ will capture the variance almost
perfectly but will be useless. To determine a reasonnable range of values for $k$ we plotted on
Supplementary Figure \ref{fig:var-vs-k} the averaged percentage of variance captured by $k$
clusters resulting from the application of k-means to the $n$ listeners. From this curve we use
the 'elbow method' to determine the range of reasonnable values for $k$. It appears that
$4= < k <= 12$ are candidates, and we looked at the average time profiles resulting from the
clustering with each value of $k$. From $k>=5$ we observed less distinct average profiles,
which is why we kept $k=4$ for the figure discussed in the main text.

\subsubsection*{Characterizing individual distributions of plays}

Supplementary Figure \ref{fig:gini-plays} gives the distribution of the listeners' Gini
coefficient resuming the heterogeneity of their distribution of plays among songs (see the
Methods section of the main text). Listeners who have a small Gini value (typically $< 0.3$)
are those whose distribution is almost flat, indicating that they listened their songs
approximately the same number of times. For such listeners the average value $P/S$ (with $P$
the total number of plays and $S$ the total number of unique songs listened) gives a clear
picture of their listening practice and of their repetition/discovery behavior. But for a large
proportion of individuals the Gini coefficient is large ($> 0.5$), revealing that these users
have concentrated their plays on a limited number of songs.

\paragraph*{S vs. P.} On Supplementary Figure \ref{fig:scaling}A we plot for each user $i$
her/his number $S_i$ of distinct songs listened versus her/his total number of plays $P_i$
(here limited to the individuals with $P < 5000$ -- which captures 95\% of listeners). Each
single point represents an individual, and we see points distributed all across the triangle
(by construction we have $S <= P$), which indicates a wide variety of profiles in terms of
repetition/exploration. Some listen to a relatively small number of songs and listen to them a
lot (small $S$ and large $P$ case), while to the opposite some other listen to many different
songs and listen each of them a few number of times (the points near the dashed line
$S=P$). Another way to capture the tendency of individuals to concentrate their plays on a
limited fraction of the songs listened is to calculate the relative dispersion $\sigma_i/\mu_i$
of their distribution of plays per song $(P_i)$. The histogram of $\sigma_i/\mu_i$ on
Supplementary Figure \ref{fig:gini-plays}B shows that there exists several types of
listeners. Individuals with $\sigma / \mu \ll 1$ have a distribution of plays per song peaked
around the mean, and the average number of plays per song $\bar{p}_i$ is hence an informative
value. To the contrary for individuals with $\sigma / \mu \gg 1$, the average number of plays
per song $P_i/S_i$ would not be representative of their listening practice.

\subsubsection*{Selection and aggregation of music genres}

Each song of a music-on-demand service is tagged with one or several tags (in the following we
name them \emph{basic genres tags} and the histogram of the number of tags per song is
displayed on Supplementary Figure \ref{fig:hist-n-basic-genres-per-song}). The weight of these
basic genres -- measured through their total number of songs in the streams -- is very
different from one genre to another. Furthermore, the analysis of the network of basic genre
tags reveal different types of tags, and some hierarchical relations between them (some
entirely include/contain others). We build the weighted and directed network $S_{ij}$ of genre
tags. It is the 1-mode projection of the bipartite network linking songs and genres tags.  In
this network the nodes represent the basic genres, and $S_{i\rightarrow j} = k$ means that
there are $k$ songs tagged with genre $i$ which are also tagged with genre $j$ (the network is
directed and in most cases $S_{i\rightarrow j} \neq S_{j\rightarrow i}$). Some basic genre tags
correspond to very broad, higher-order categories (e.g. 'Pop', 'Rock' 'Alternative', 'Dance')
and serve as coarse-grained classifiers. The analysis of this network reveals that it has a
hierarchical structure and that some of these genres tags entirely ``contain'' smaller, more
informative tags. For example all ``Blues'' songs are also tagged as being ``Rock'' songs, and
all ``Metal/Hard-Rock'' songs are also tagged as ``Rock''. On the contrary, there are some
``Rock'' songs that are tagged only as ``Rock''. The purpose of the filtering we performed and
detail below was then to keep the most informative/precise tag(s) for each song, whenever it is
possible and relevant.

To perform our genre analysis, we start with the same dataset $D$ used in subsections 1
(Rhythms), 2 (Difference and repetition) and 4 (Playing music at a party) of the results
section in the main text. This dataset contains the entire streams history of the 4,615 users
during 100 days.  We first merge this dataset with the songs database provided by the
music-on-demand company, which contains the genre tags associated to each song of the
catalog. It results in a dataset $D'$ giving us the complete streams associated to each of the
basic genre tags. We filter this dataset of $\approx 2*10^7$ entries by applying the following
rules:

\begin{itemize}

\item we remove the purely ``geographical'' tags (\emph{World, France, Europe, North
      America, Central America/Caribean, South America, Brazil, Africa, Maghreb, Middle East,
      Australia/Pacific}) which give limited information on the music genre itself;

\item we filter out the basic genre tags which account for less than $0.01\%$ (nb: arbitrary
  parameter choice) of the songs listened -- \emph{including "Bollywood", "Finnish folk",
      "Medieval", "Chaabi", "Comptines/Chansons", "Mento/Calypso", "K-Pop", "Celtic music,
      "Bachata", "Classical turkish music'', "Axé/Forró", "Regional méxicain", "Instrumental
      Hip Hop, "Mariachi", "Argentinian folklore'', "German rap", "Banda", "Nederlandstalige
      volksmuziek", "Brasilian rock", 
      "South-African House","Teen thaï", etc.} --, in order to reduce the number of genres
  compared and focus on the most listened ones; after this step we have discarded
  $\approx 40$\% of the basic genre tags.

\item we keep all the streams of songs which are tagged with \emph{one basic genre tag only}
  ($\approx 3\times 10^5$ songs); the basic genres tags that are used as single tags are the
  following: \emph{Hip Hop, Dance, Pop, R\&B/Soul/Funk, Reggae, Electro, Rock, Alternative,
    International pop, Jazz, Classical, Variété, Country, Brasil, Music for kids, French
    chanson, Movies/Video games, Tropical, Latin rock}.

\item for the remaining streams of songs tagged with two or more basic genres tags
  ($\approx 2.5\times 10^5$ songs), we inspect the statistic of the pairs of \emph{basic genres
    tags} among songs. For each pair of basic genres $(g,g')$, if it appears that $g$ entirely
  contains $g'$ (i.e. all songs tagged as $g'$ are also tagged as $g$), then we remove $g$ from
  the corresponding streams (i.e.  streams which were previously associated to these two basic
  genres will now be associated only with genre $g'$). For example, let's say we wish to
  calculate aggregated statistics for ``Rock'', ``Blues'' and ``Metal'' streams. Since all
  songs tagged as ``Blues'' (resp.  ``Metal'') are also tagged as ``Rock'', to come up with
  more relevant statistics for the three genres we consider that it is more significant to
  discard streams associated to ``Blues'' and ``Metal'' songs when computing reference
  statistics for ``Rock'' streams (and keep in the ``Rock'' category limited to the songs
  tagged solely as ``Rock'' ). We end up in removing basic genre tags such as \emph{Pop, Rock,
    Electro, Hip Hop, Jazz, Classical, etc.} among the tags that qualify songs with 2 or more
  tags, because they are systematically associated with more informative tags (e.g. \emph{indie
    rock, Metal/hard-rock, Blues, Techno/House, rock'n roll/rockabilly, Funk, chill
    out/trip-hop, instrumental jazz, instrumental hip-hop, opera, etc.}).

\end{itemize}

We end up with more than $7\times 10^6$ streams.  The statistics of the number of basic genre
tags per song in the remaining filtered streams are given in Table
\ref{table:hist-n-genres-songs}. More than 90\% of the songs are tagged with 1 genre only, and
almost all songs (98\%) are tagged with 1 or 2 genres, limiting the risks of confusion in the
analysis of listening patterns associated to various genres.

\begin{table}
  \centering
  \begin{tabular}{lll}
    \hline
    \#genres & \#songs & weight \\
    \hline
    1 & 439582 & 9.002220e-01\\
    2 & 38348 & 7.853304e-02\\
    3 &  9384 & 1.921754e-02\\
    4 & 970 & 1.986467e-03\\
    5 & 20 & 4.095809e-05\\
    \hline
  \end{tabular}
  \caption{\label{table:hist-n-genres-songs} Number and proportion of songs having a given
    number $k$ of genre tags, after filtering and removing higher-order genre tags.}
\end{table}

\onecolumngrid
\newpage
\clearpage 

\section*{Supplementary Figures}

\vspace*{3cm}
\begin{figure}[!htbp]
  \centering
  \includegraphics{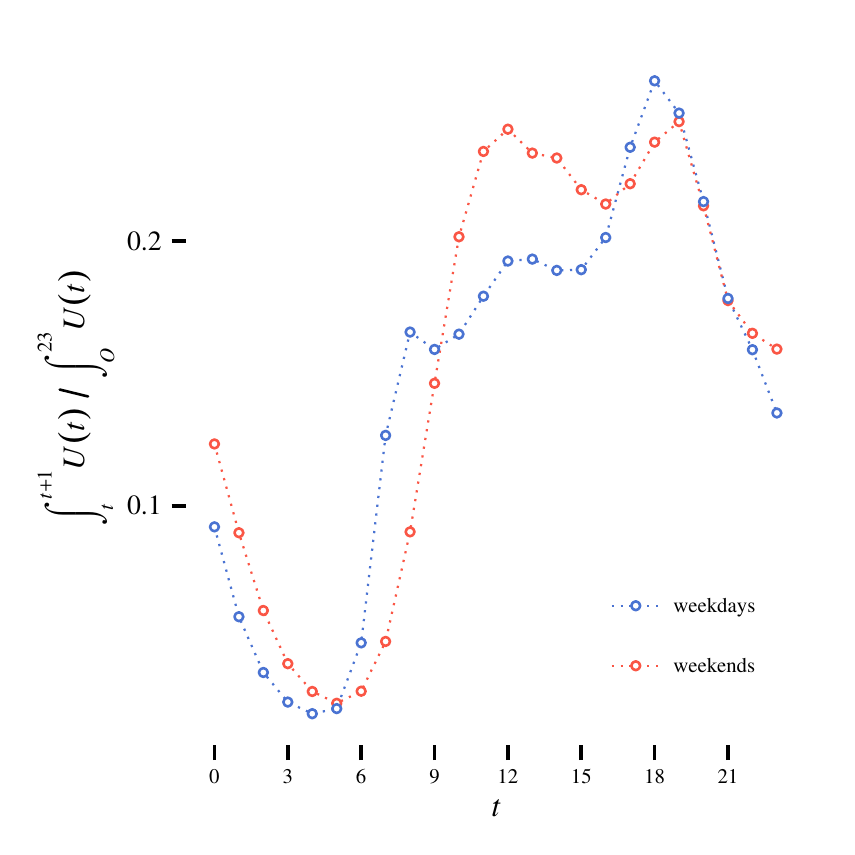}
  \caption{ \label{fig:plays-users-vs-time}\textbf{Hourly evolution of the proportion of active
      listeners, during an average weekday and an average weekend day.}During weekdays the
    number of individuals listening to music increases over the day to reach a maximum around
    7-8 p.m. We notice two small peaks, one in the morning aroung 8 a.m. and the other one at
    lunchtime around 1 p.m. Most of the plays occur during the afternoon and evening. During
    weekends the collective rhythm is different with two peaks, one around 12 p.m. and the
    other in the early evening, around 7-8 p.m. (error bars are small indicating a high
    regularity of temporal listening habits}
\end{figure}

\newpage

\begin{figure}
  \centering
  \includegraphics{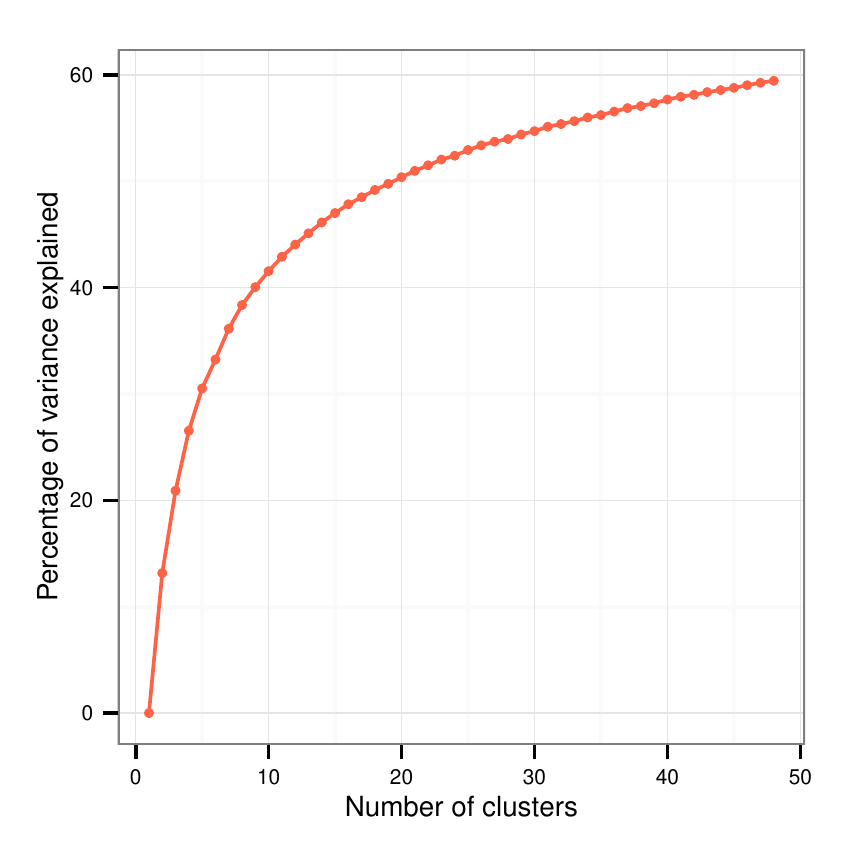}
  \caption{ \label{fig:var-vs-k} Percentage of variance explained when clustering listeners in
    $k$ groups according to their listening time profile average over 100 days.}
\end{figure}

\newpage

\begin{figure}
  \centering
  \begin{tabular}{cc}
    (a) & (b) \\
    \includegraphics{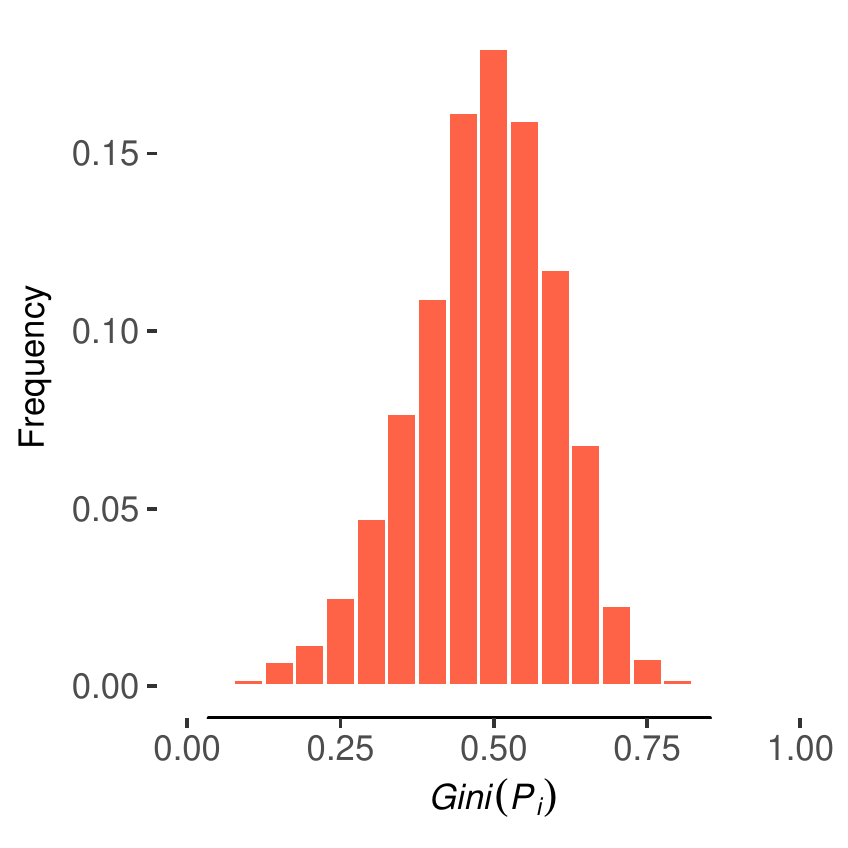} &
    \includegraphics{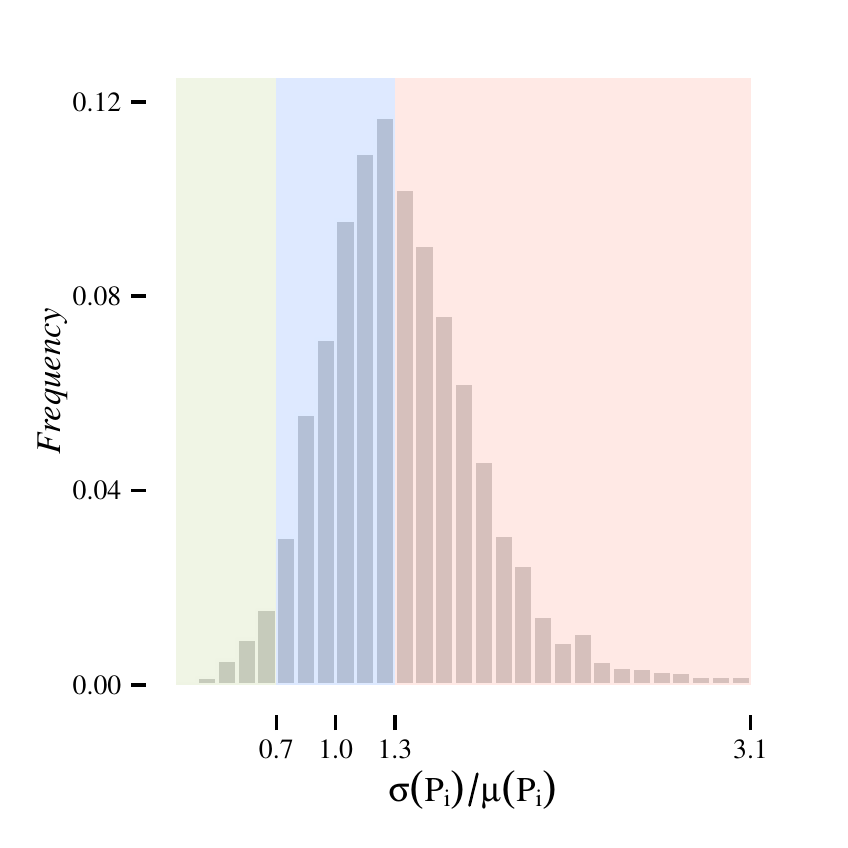} 
  \end{tabular}
  \caption{ \label{fig:gini-plays}(a) Histogram of the individuals' Gini normalized coefficient
    resuming their distribution of plays among the songs they listened (b) Histogram of the
    users' relative dispersion of plays. For each user $i$, $\mu_i$ is the mean value of
    his/her distribution of plays per song, while $\sigma_i$ is the standard deviation of this
    distribution.}
\end{figure}

\newpage

\begin{figure}
  \centering
    \begin{tabular}{ccc}
      (a) & (b) & (c) \\
      \includegraphics[width=2in]{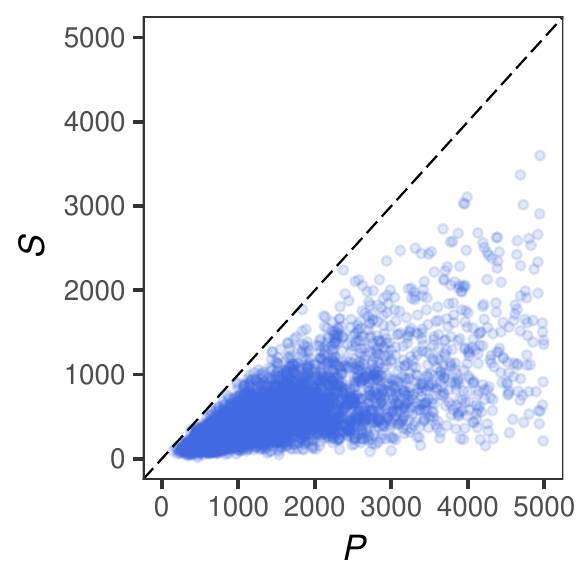} &
  \includegraphics[width=2in]{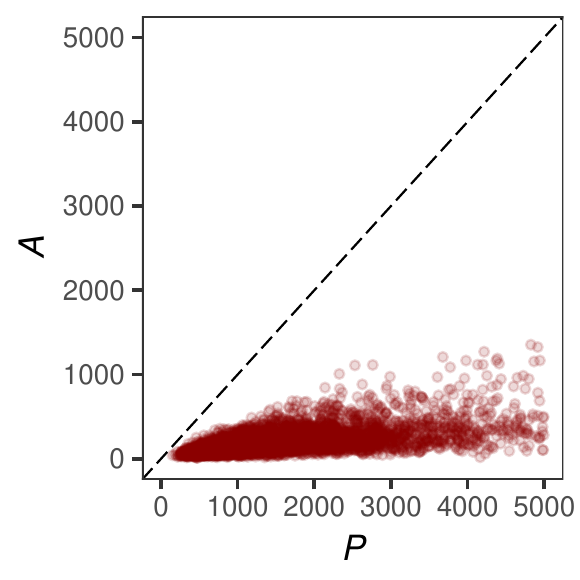} &
  \includegraphics[width=2in]{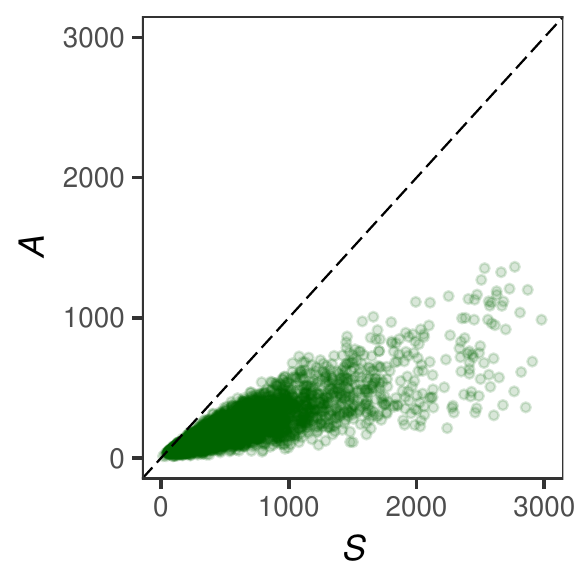}
    \end{tabular}                                                                        
    \caption{ \label{fig:scaling} On these scatterplots each dot represents a listener (a)
      Number of unique songs listened $S$ vs. total number of plays $P$ (b) Number of unique
      artists listened $A$ vs. total number of plays $P$ (c) $A$ vs $S$. The data appears
      clouded, we see a lot of fluctuations and no clear relation linking $P$, $S$ and $A$ that
      would allow to make predictions.}
\end{figure}

\newpage

\begin{figure}
  \centering
  \includegraphics[width=3.42in]{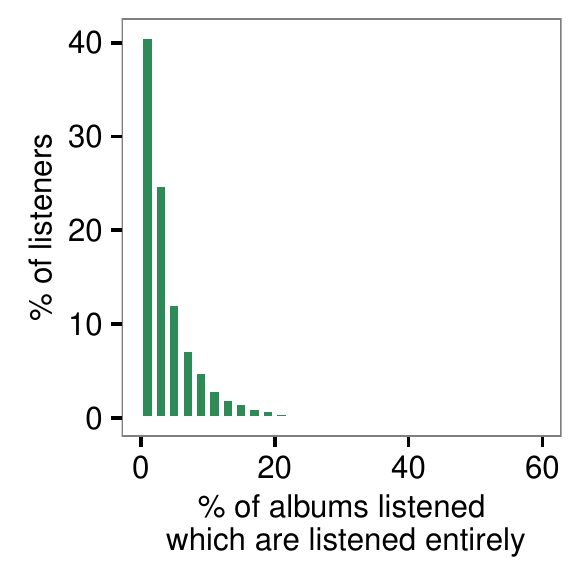}
  \caption{\label{fig:hist-albums-entirely-listened} \textbf{Proportion of albums listened
      entirely (but not necessarily in just one time and sequential order).} More than 50\% of
    the users listened entirely less than 5\% of all the albums in which they grabbed songs,
    underlining that album-oriented listening sessions are clearly not the standard practice on
    streaming platforms.}
\end{figure}

\newpage

\begin{figure}
  \centering
  \includegraphics{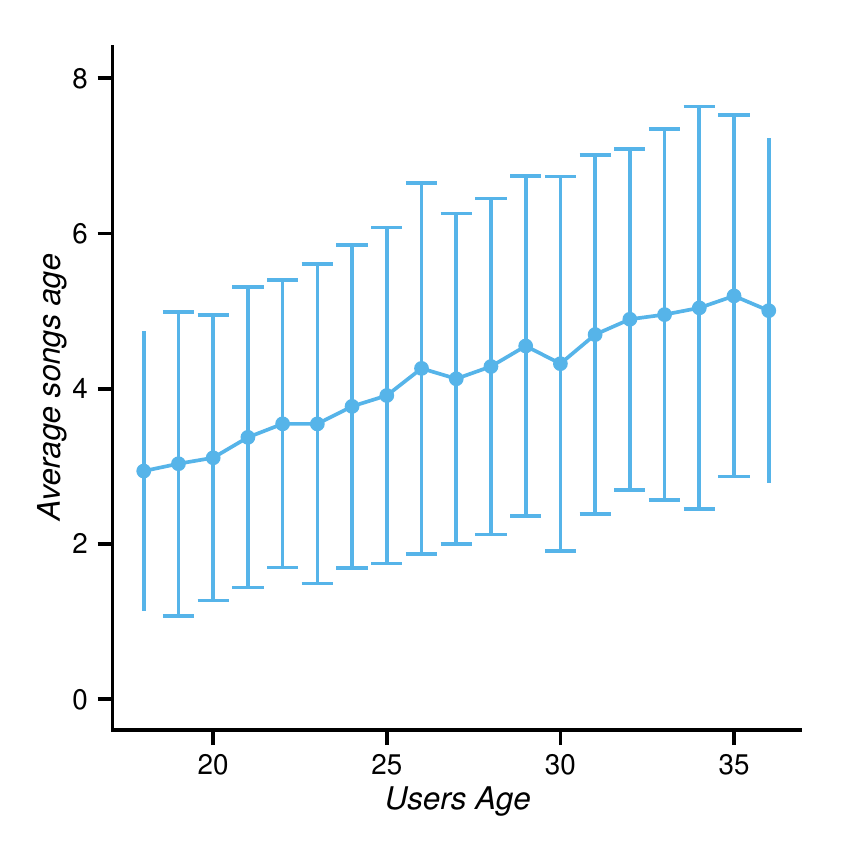}
  \caption{\label{fig:songs-age-vs-users-age} \textbf{Do we stay true to the music of our
      generation?} Average age of songs listened as a function of the listener's age. While
    there are of course a lot of fluctuations and different listening behaviours inside a given
    'demographic group', we observe a tendency: as people get older, they tend to listen to
    older music as well.}
\end{figure}

\pagebreak


\begin{figure*}
  \centering
  \includegraphics{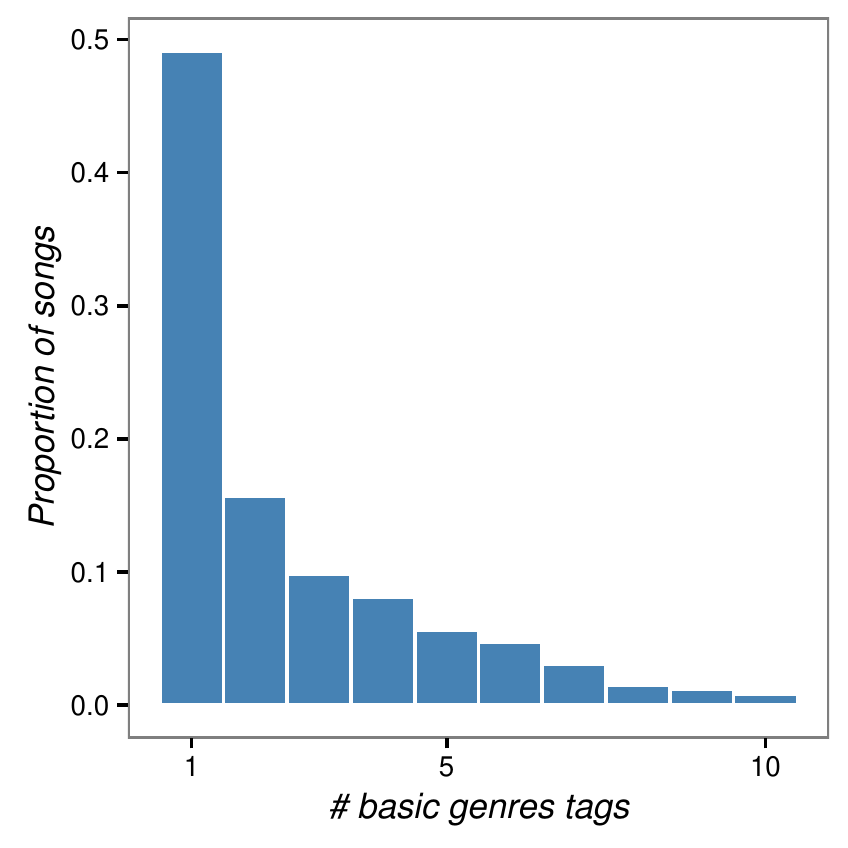}
  \caption{ \label{fig:hist-n-basic-genres-per-song}\textbf{Histogram of the number of basic
      genre tags per song, before filtering.} The songs considered for building the histogram
    are those which have been listened to during the 100 days period under scrutinity. }
\end{figure*}

\newpage

\vspace*{3cm}
\begin{figure}
  \centering
  \includegraphics[width=\textwidth]{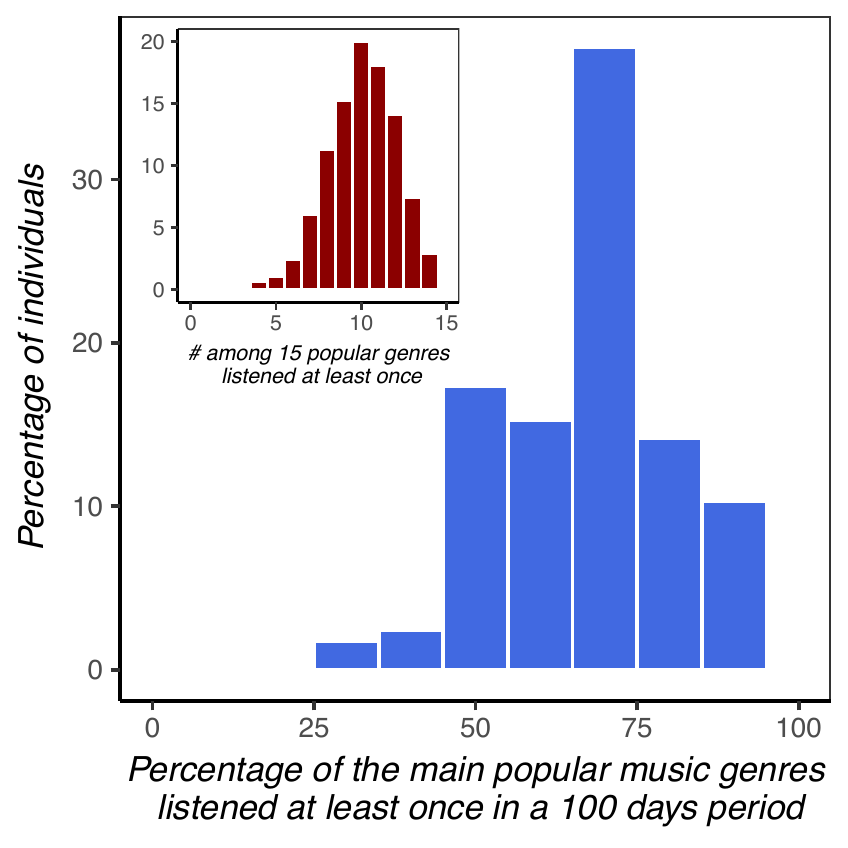}
  \caption{\label{fig:eclectism-at-first-sight}\textbf{Listeners seem eclectic at first
      sight.} Histogram of the proportion of the main genres listened at least once in a 100
    days period ; (inset) Histogram of the number of genres listened at least once in a 100
    days period among 15 classic popular broad music genres (\emph{Pop, Rock, Hip Hop, Dance,
      R\&B/Soul/Funk, Reggae, Electro, Alternative, Country, Jazz, Classical music, Chanson,
      'Variété', Tropical}).}
\end{figure}

\newpage

\vspace*{3cm}
\begin{figure*}
  \centering
  \includegraphics[width=\textwidth]{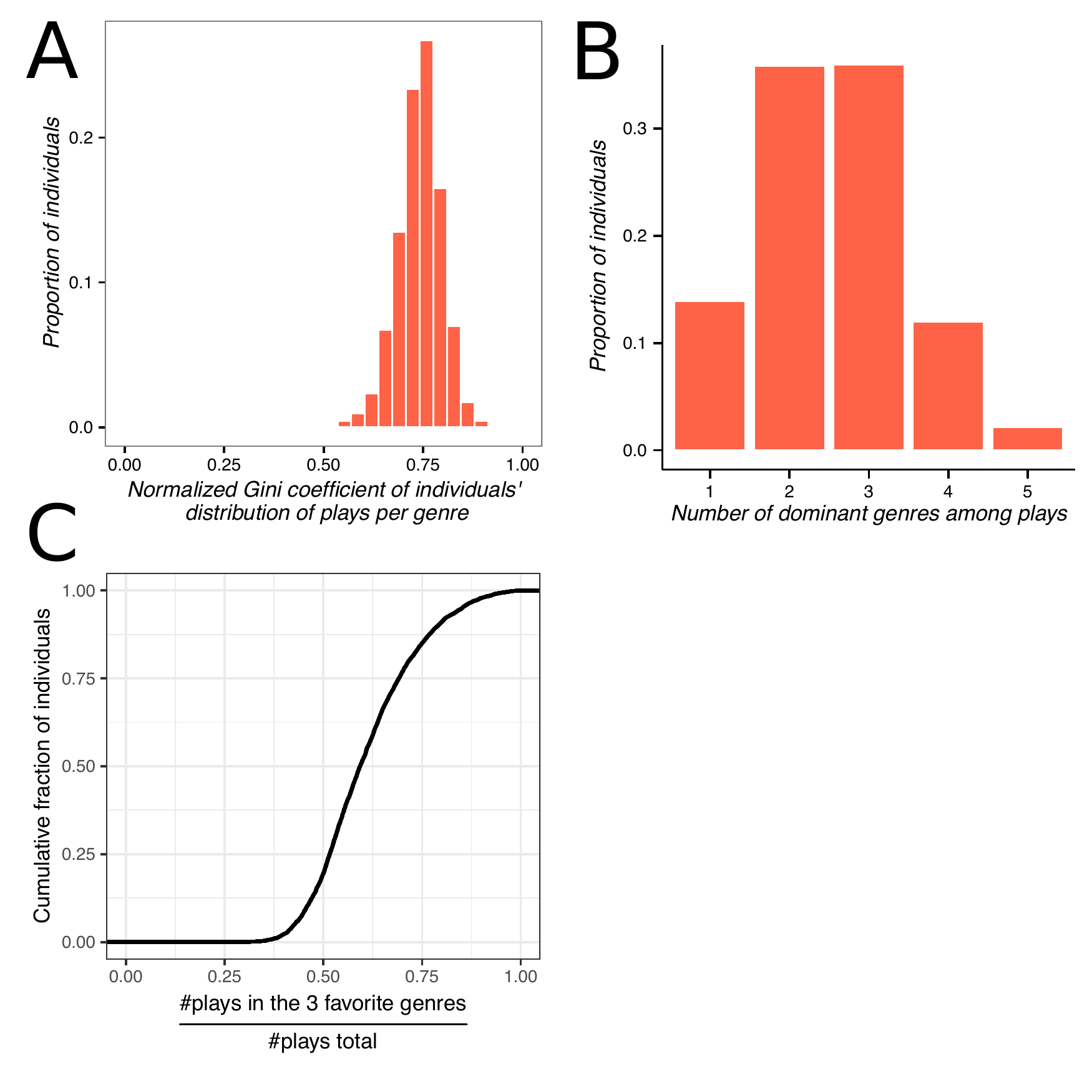}
  \caption{ \label{fig:genres-repartition} \textbf{The listeners' inegal repartition of plays
      among the music genres they listen to.}  (a) Distribution of the Gini coefficient values
    of individuals that resume the inequality of their distribution of plays among the
    different main music genres (b) Histogram of the number of dominant genres $D$ among
    individuals (see Methods in the main text for details about the calculation of the number
    of dominant terms in a distribution from its Gini coefficient) (c) CDF of the weight of the
    3 favorite genres (determined from the number of plays) among listeners.}
\end{figure*}

\newpage

\vspace*{3cm}
\begin{figure*}
  \centering
  \includegraphics[width=\textwidth]{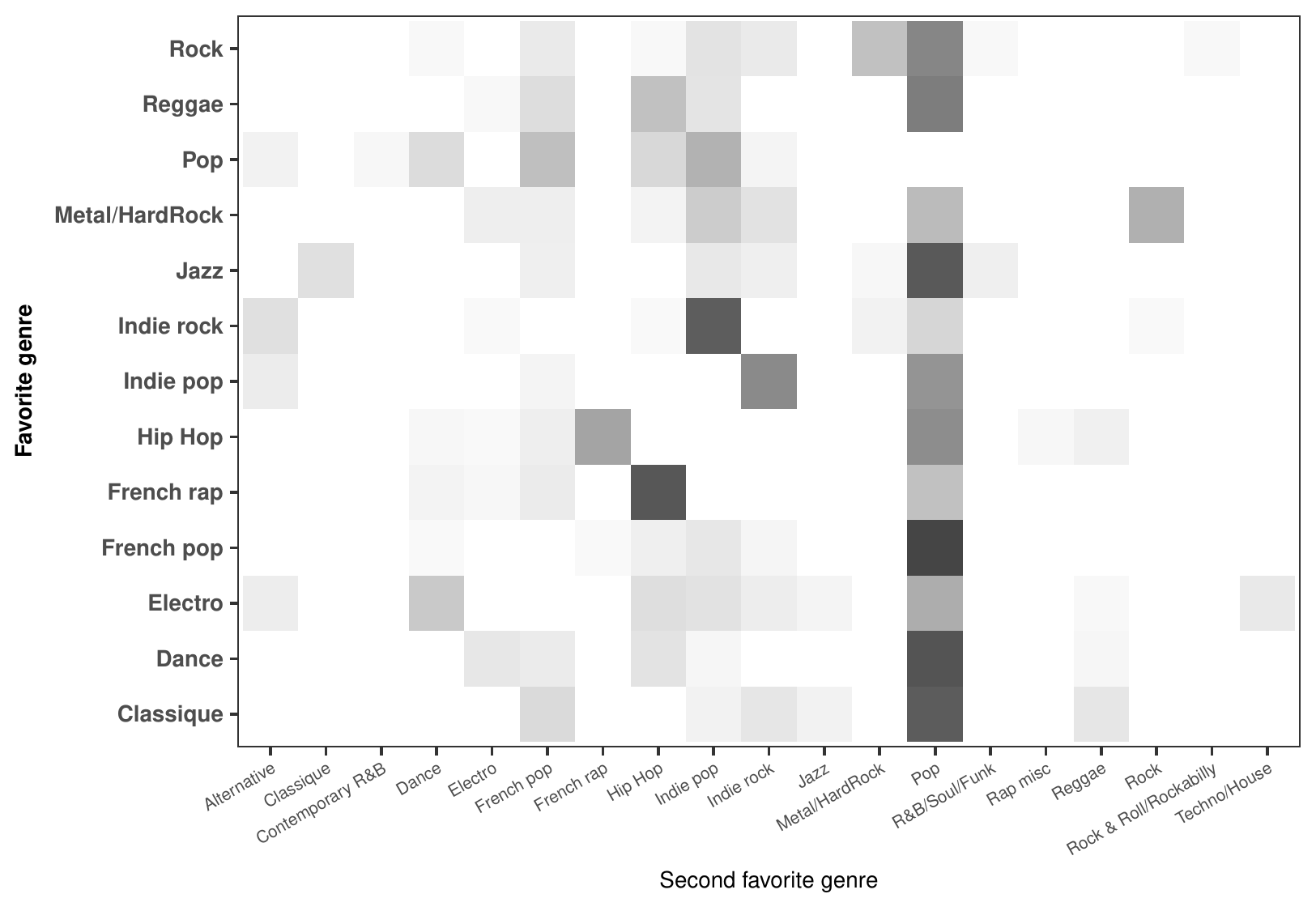}
  \caption{ \label{fig:favorite-genres}\textbf{Conditional probabilities to observe pairs of
      favorite genres among listeners.}  The color intensity of each cell is indexed on the
    probability $p(g_2/g_1)$ that a listener having $g_1$ as favorite genre has $g_2$ as second
    favorite genre.}
\end{figure*}

\newpage

\vspace*{2cm}
\begin{figure}
  \centering
  \includegraphics[width=\textwidth]{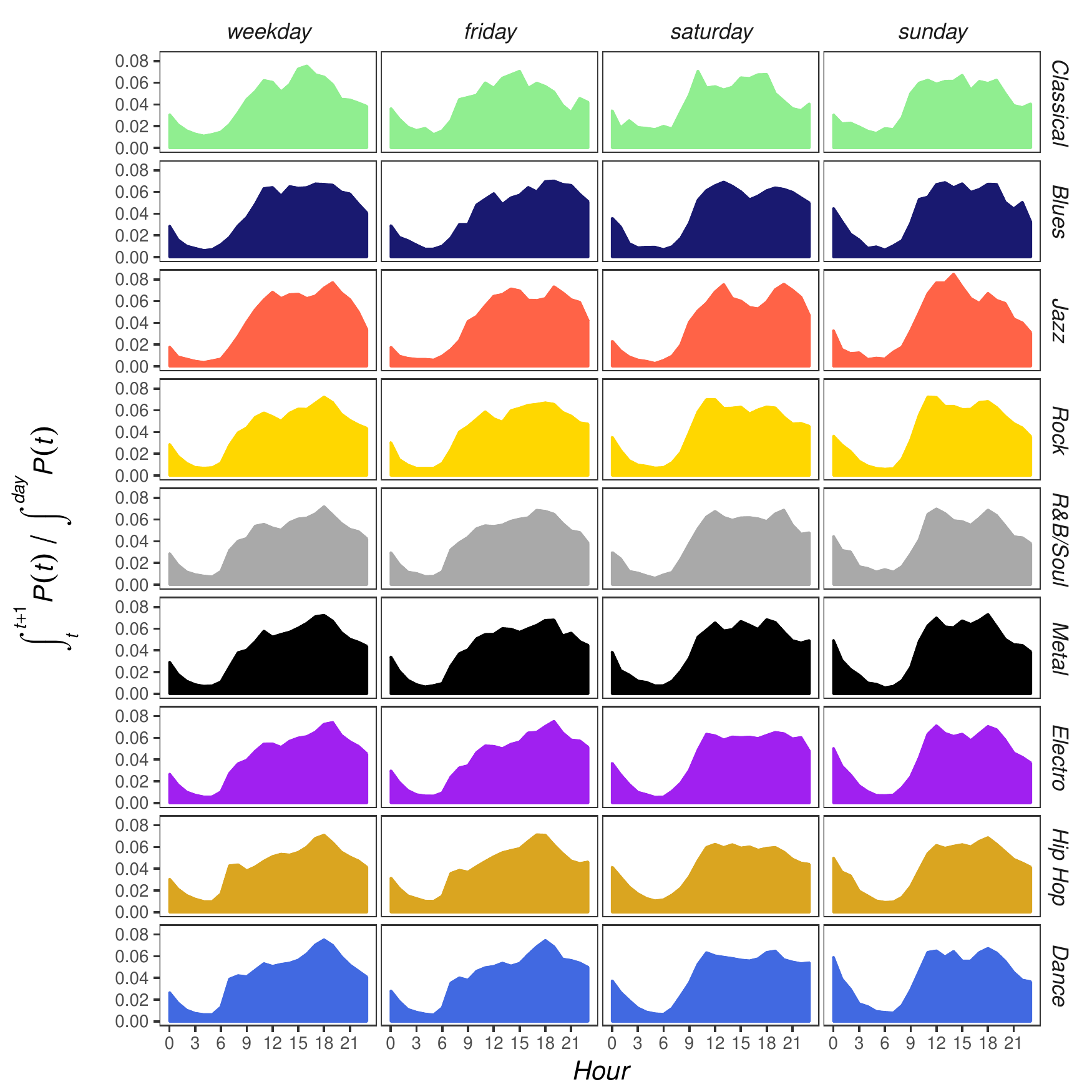}
  \caption{ \label{fig:genres-plays-vs-time} \textbf{Relative proportion of plays as a function
      of the hour of the day and day of the week, for different music genres.} Let alone micro
    variations -- which might be due to the relatively small number of listeners of certain
    genres at certain hours (nb: the total number of listeners under study is 4615) and to
    statistical fluctuations -- we see no significative differences between the average
    listening time profiles of the genres.}
\end{figure}

\newpage

\vspace*{5cm}
\begin{figure*}
  \centering
  \includegraphics[width=\textwidth]{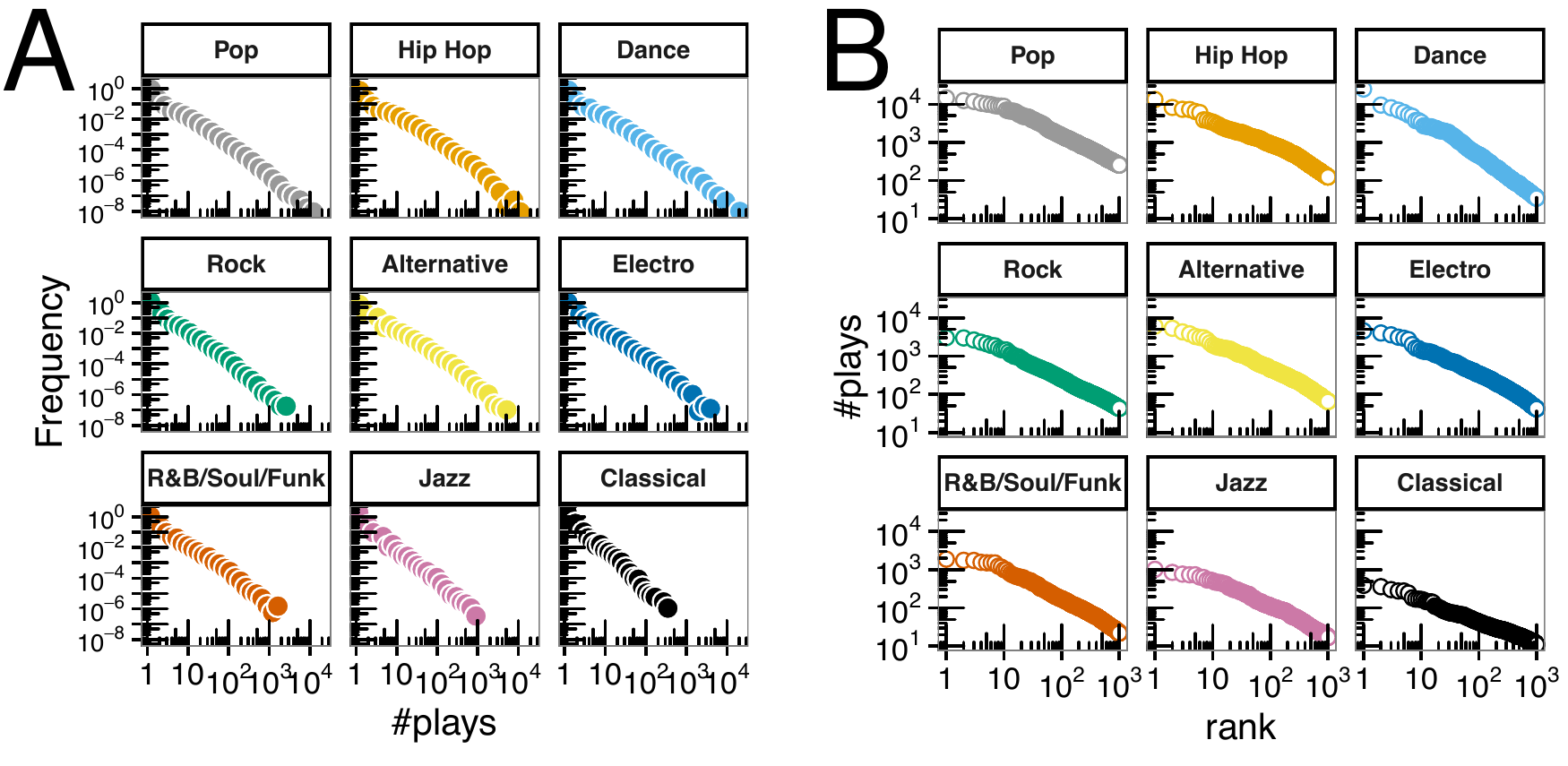}
  \caption{\label{fig:genres-distr}\textbf{The hierarchy of songs in various genres.} (a)
    Histogram of the number of plays per song for the same set of genres (b) Rank-size plot for
    the 1000 most popular songs in the streams of nine classic music genres.}
\end{figure*}

\newpage

\vspace*{3cm}
\begin{figure}
  \centering
  \includegraphics[width=\textwidth]{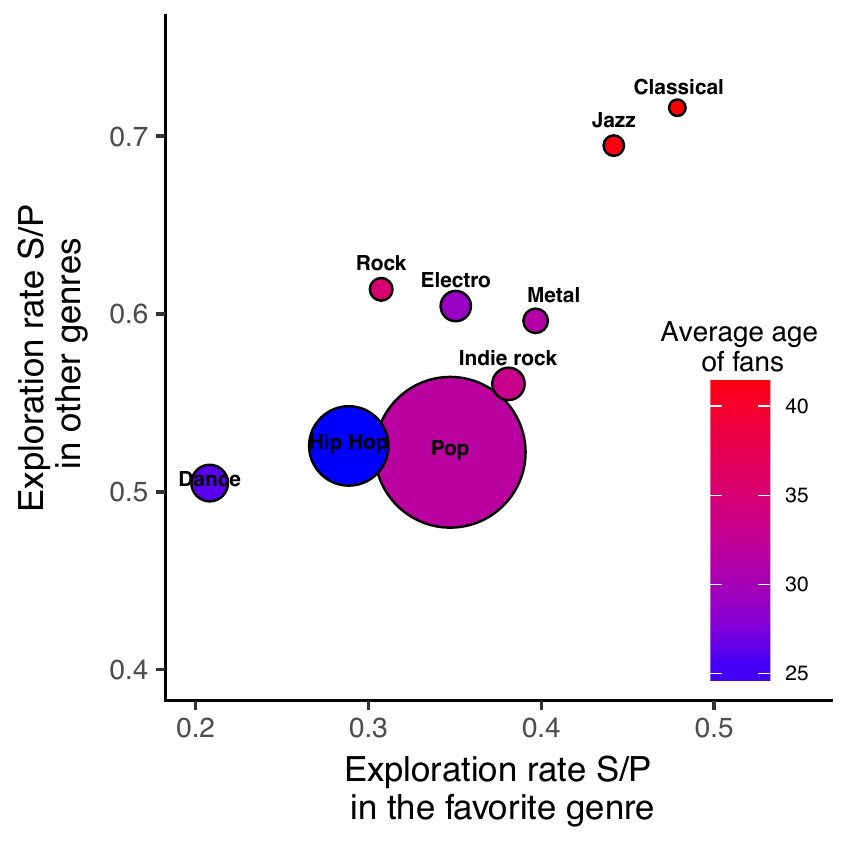}
  \caption{\label{fig:fans-statistics} Exploration statistics for groups of fans of various music
    genres, along with the average age of the group members (fill colour) and its importance among the
    population of listeners (circle size).}
\end{figure}

\newpage

\end{document}